\documentclass[12pt]{article}
\usepackage[utf8]{inputenc}
\usepackage{url}
\usepackage{subfigure}
\usepackage{amsmath}
\usepackage{amssymb}
\usepackage{xcolor}
\usepackage{graphicx}
\usepackage{enumerate}
\usepackage{etoolbox}
\usepackage[left=2.5cm,right=2.5cm,top=2.0cm,bottom=3.cm]{geometry}

\usepackage[style=ieee,url=false,backend=biber]{biblatex}
\usepackage{hyperref}

\title{Transformers for scientific data: a pedagogical review for astronomers}
\author{Dimitrios Tanoglidis, Bhuvnesh Jain, Helen Qu \\
University of Pennsylvania}
\date{\vspace{-0.3in}}

\begin{document}

\maketitle

\begin{abstract}
    The deep learning architecture associated with ChatGPT and related generative AI products is known as a transformer. Initially applied to Natural Language Processing, transformers and the self-attention mechanism they exploit have gained widespread interest across the natural sciences. The goal of this pedagogical and informal review is to introduce transformers to scientists. The review includes the mathematics underlying the attention mechanism, a description of the original transformer architecture, and a section on applications to time series and imaging data in astronomy. We include a Frequently Asked Questions section for readers who are curious about generative AI or interested in getting started with transformers for their research problem. 
\end{abstract}

\section{Introduction}

This is a short, pedagogical set of notes on the \textit{attention mechanism} and the deep learning architecture that utilizes this mechanism, \textit{transformers}. Transformers have revolutionized Natural Language Processing (NLP) and, more recently, have also emerged as a powerful paradigm for computer vision applications (Vision Transformers - ViTs) and, more generally, imaging or multi-modal datasets in the sciences. Transformer-based architectures are behind most of the AI applications that have made it to the news recently, including Large Language Models (LLMs) like GPT and image-from-text generators like DALL-E. The goal of this short review is to introduce scientists, astronomers in particular, to transformers. 

These notes are intended to be a quick start to mathematically oriented scientists. Our focus is on understanding the main concepts (and math) behind the self-attention mechanism and the transformer block, which is the fundamental building block for transformer-based architectures. We will present the key equations underlying the attention mechanism, but will \textit{not}  describe any complete architecture or model (such as the Large Language Models  BERT or GPT) or how they are trained. Furthermore, we start our exploration \textit{in medias res}, showing the modern self-attention mechanism and omitting the historical introduction about the attention mechanism as part of a Recurrent Neural Network model. We also do not cover variations of the attention mechanism that have been developed for a variety of applications.

The key property of self-attention is that it allows transformers to capture long-range dependencies without sequential processing, parallelizing the training process and simultaneously avoiding vanishing gradient problems. This means that for any sequential data setting with a large body of training data available for self-supervised learning, transformers are a natural choice. They have been shown to  outperform alternatives, in particular Recurrent Neural Networks (RNNs) in NLP. As GPT has demonstrated, with sufficiently large training data, transformer architectures also show remarkable ``emergent'' properties, or  abilities in large models that are not present in smaller models. These include the ability to perform a new task with little to no task-specific training data (few- and zero-shot learning) and learning a new task from a single prompt, which does not alter the network weights (in-context learning)\footnote{See https://arxiv.org/abs/2206.07682 for a review on emergent properties of large language models.}. 

The physical sciences are increasingly applying transformers to solve data-rich problems. In astronomy, time series data is common in subfields spanning variable stars, supernova cosmology, quasars, gamma-ray bursts and moving objects in the solar system. With data challenges based on simulated data (e.g. 
PLAsTiCC for the forthcoming LSST survey), a large body of available observational data (e.g. the recently completed ZTF survey), and the transformative time domain Rubin-LSST   survey scheduled to begin by 2026, the field is ripe for new machine learning approaches. Transformers are therefore a natural choice for time series problems as described further below. 

Static datasets of images, spectra and other multi-wavelength observations can be analyzed using variations such as Vision Transformers. The success of transformer-based architectures in tackling heterogeneous data (e.g. Microsoft's climate modeling package) and multi-modal data is appealing as well, since such datasets in astronomy and other sciences are often under-utilized due to the technical challenges of synthesising data of diverse types and sources. We will focus on applications of ViTs to image analysis in these notes. 

We begin with a mathematical account of the self-attention mechanism in Section 2, including brief explanations of multi-head attention and positional encoding. The transformer block, the basic element of a transformer-based deep learning architecture, is described in Section 3. Two key applications to astronomy, time series and image analysis, are summarized in Section 4, and we conclude with a set of Frequently Asked Questions in Section 5.  We  refer the interested reader to some excellent tutorials and papers in Sec. \ref{sec: References} and to our FAQ below for a more complete understanding of the attention mechanism, its historical evolution, alternatives, and different transformer-based architectures.

\section{The self-attention mechanism}


\subsection{Basic form of  self-attention }

The most important mechanism behind any modern transformer is the \textbf{self-attention} mechanism presented in Vaswani et al (2017). We will start by describing a simple version of it that clearly illustrates the basic idea behind it, and then we can generalize to present the form of self-attention as it is actually used in modern transformer architectures.

Transformers, and the attention mechanism they are based on, are mainly a way to convert one set of \textbf{input embeddings} to another set of \textbf{output embeddings, or representations}. An \textit{embedding} is nothing but a vector that represents a piece of data. Embeddings are most famously applied in NLP, since words have to be converted to a numerical representation to be used in machine learning algorithms. Efficient algorithms like Word2Vec \cite{Mikolov_2013} demonstrated that it is possible to create vector representations of words that capture linguistic regularities, such as the concept of gender in the famous example: $\vec{king} - \vec{male} + \vec{female} \cong \vec{queen}$\cite{mikolov2013linguistic}.

Since attention converts a set of input embeddings to another set of embeddings\footnote{We will use the term \textit{embedding} only for the first conversion of raw input data to a vector, and the term \textit{representation} for subsequent transformations described in the next sub-section}, a powerful way to understand the inner workings of the attention mechanism is to keep track of how the dimensions of a sequence of input vectors change as they progress through the mechanism's operations.

Let's suppose we have an input sequence of $T$ vectors, each of dimension $d_e$ (the input embedding dimension): \textit{vectors} $\mathbf{x}_1,\mathbf{x}_2,\dots,\mathbf{x}_T$ (collectively they can be represented by a matrix $\mathbf{X}$ with dimensions $T \times d_e$). These vectors can represent time series data, a (flattened) part of an image, or as in the case of LLMs ``tokens" that represent words, parts of words, punctuations etc. These vector representations (\textit{embeddings}, as described earlier), are generally learned from the raw input data. 

The point of the embedding is that \textit{related inputs are represented by similar vectors (defined by e.g., cosine similarity)}. The  self-attention mechanism  \textbf{transforms} the input sequence to an output sequence, $\textbf{y}_1,\textbf{y}_2,\dots,\textbf{y}_T$, where each element of the output is a weighted sum of the input sequence: 
\begin{equation}
    \mathbf{y}_i = \sum_j W_{ij} \mathbf{x}_j.
\end{equation}
This goal of the transformation is to create \textit{context-aware} output  vectors, i.e. output vectors that account for the pair-wise relationships between input vectors. Why do we want that? An example from the NLP domain may make it clear. Consider the following two sentences: ``The Queen met with delegates from the member states.", and ``Gary Kasparov used the Queen to start his attack against Deep Blue in the first match held in Philadelphia". It is clear that the word ``Queen" in the first case refers to the head of state, while in the second case to the chess piece. However, in the original embedding, the word queen is represented by a single vector that cannot convey these two different meanings. The above transformation aims to create representations that are aware of the surrounding words, and thus the context the word is used in (politics vs. chess). (Note that we use the term \textit{embedding} vectors for the input vectors $\mathbf{x}_i$ and \textit{representations} once they are acted on by the weights, though the latter are sometimes also  called embedding vectors. )

In the attention mechanism, the projection matrix $W_{ij}$ (also called ``weight matrix'' as its elements are weights) is derived from the inputs; specifically, we want to give higher weight to those inputs that have a higher similarity to $\textbf{x}_i$. Since the dot product is a measure of similarity, we can define:
\begin{equation}
    w_{ij} = \textbf{x}_i^T\textbf{x}_j.
\end{equation}
To convert the above dot product to weights that are in the range [0,1] and their sum equals  1,  we pass this result through a softmax function to generate the final weights $W_{ij}$ (called \textit{attention weights}):

\begin{equation}
    W_{ij} = \frac{\exp w_{ij}}{\sum_j \exp w_{ij}} = \mbox{softmax}(w_{ij})
\end{equation}

That's it! This is the core of the attention mechanism. The initial sequence of vectors has been \textbf{transformed} into a different one, a weighted combination of all the other vectors, with weights being larger for input vectors that are more similar (it ``attends'' more to them). 

\subsection{Adding learnable weights: query, key, and value parameterization}
\label{Sec: QKV_parametrization}

We haven't discussed why the above method of transforming a sequence can be beneficial in machine learning tasks. This can be better understood once we discuss one more layer of complexity. 

The first thing to notice in the above formulation is that we have not introduced any learnable parameters; the weights $W_{ij}$ are solely determined by the input vectors. A model based on that basic operation cannot be optimized for a specific task (e.g. classification) via the standard machine learning methods.

To make progress, note that in the above formulation, every vector $\textbf{x}_i$ is used three times in three different roles: we can write the sum explicitly as $y_i = \sum_j \mbox{softmax}(x_i^Tx_j)x_j$. This allows us to modify each occurrence of the vector $\mathbf{x}_i$ by multiplying it with a different 
matrix containing \textit{learnable} weights. These matrices,  $\mathbf{W}_Q$, $\mathbf{W}_K$, $\mathbf{W}_W$,  can each be multiplied with the input vector to give three new vectors, called Query, Key, and Value, which are given as follows (using row vector notation):

\begin{eqnarray}
\mbox{Query: } \mathbf{q}_i = \mathbf{x}_i \mathbf{W}_Q \\
\mbox{Key: } \mathbf{k}_i =  \mathbf{x}_i\mathbf{W}_K \\
\mbox{Value: } \mathbf{v}_i =  \mathbf{x}_i\mathbf{W}_V 
\end{eqnarray}

These names may not seem intuitive at first. Let us make an analogy with a YouTube video search\footnote{Thanks to MIT's intro to deep learning course for this example. Also, note that the description provided here is an extremely simplified version of how YouTube search algorithm works.}. Assume that you want to find a YouTube video on how to bake a pie. You will write on the search tab ``How to bake a pie". This is the Query. Now, every YouTube video has some keywords related to its contents (e.g. ``baking pie", ``cute cat", ``quantum mechanics" etc). These are the Key values associated with every video containing content information. When searching for a relevant video, the similarity of the Query to the Keys of each video is computed, and the videos where this similarity score is the highest are returned at the top of our search. The content of these top-ranked videos contains the Value (information) that best answers our initial question (Query). Similarly, in the attention mechanism, each input vector acts as a Query that is compared to every other vector (Keys) to find those combinations with the highest similarity (the attention score weights $W_{ij}$). The output is a linear combination of the input, dominated by those input vectors (Values) with the highest attention scores.


Let us now return to the mathematical details of the attention mechanism in this more flexible Query--Key--Value parameterization. The matrices $\mathbf{W}_Q$, $\mathbf{W}_K$, $\mathbf{W}_V$ have dimensions $d_e \times d_q$, $d_e \times d_k$, and $d_e\times d_v$, respectively; they are \textit{projection} matrices that project the vectors $\textbf{x}_i$ into the lower-dimensional query, key, and value spaces. The projection dimensions $d_q=d_k$;  $d_v$, can in principle differ from them, but in practice is often chosen to be the same. 

The attention weights $W_{ij}$ can now be calculated via the \textbf{scaled dot-product} between the query and key vectors:  
\begin{equation}
W_{ij} = \mbox{softmax}\left( \frac{\textbf{q}_i\textbf{k}_j^T}{\sqrt{d_k}}\right)
\end{equation}
The scaling by $\sqrt{d_k}$ ensures that the range of the argument to the softmax does not increase with the number of dimensions. This  in turn helps the stability of gradient calculations during training since the softmax function is sensitive to large values of the argument. 

\begin{figure}[!ht]
\centering
\includegraphics[width=0.45\textwidth]{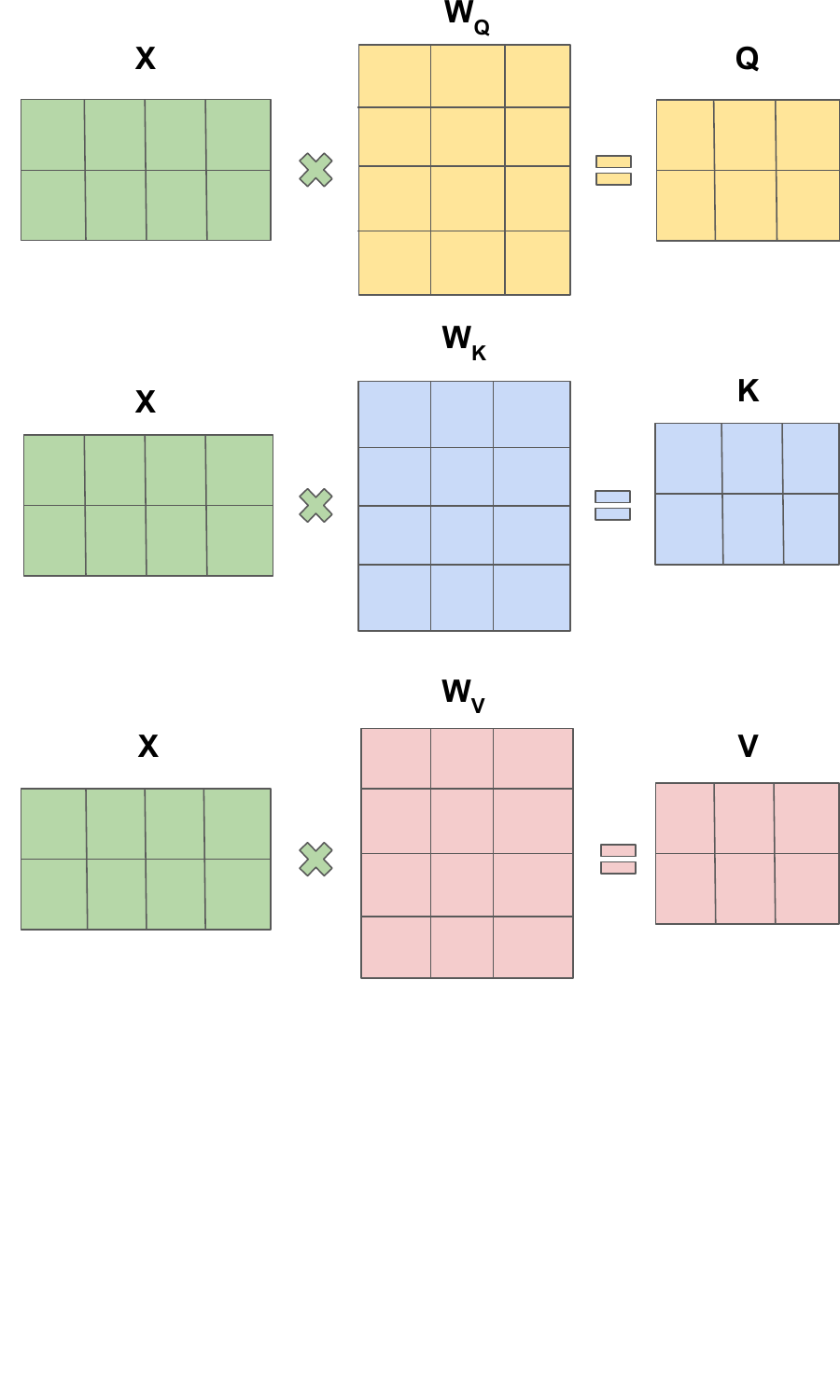}
\vspace{-2.8cm}
\caption{The matrix containing the input embeddings ($\mathbf{X})$ is multiplied with the three projection  matrices ($\mathbf{W}_Q$,$\mathbf{W}_K$,$\mathbf{W}_V$), to produce the query, key, and value representations  ($\mathbf{Q}$,$\mathbf{K}$, $\mathbf{V}$). Adapted with modifications from \url{http://jalammar.github.io/illustrated-transformer} }
\label{Fig: Linear_Projection}
\end{figure}

Finally, the output sequence is given by the following weighted sum of values:
\begin{equation}
    \mathbf{y}_i = \sum_j W_{ij} \mathbf{v}_j.
\end{equation}

In practice, we  define matrices $\mathbf{Q}, \mathbf{K}, \mathbf{V},$ whose $i-$th rows are the vectors $\mathbf{q}_i,\mathbf{k}_i,\mathbf{v}_i$, respectively. They have dimensions $T \times d_q, T\times d_k,$ and $T\times d_v$, respectively. 
Fig. \ref{Fig: Linear_Projection} shows how these matrices are created from the input and the projection matrices. Then, the \textbf{scaled dot-product} attention can be expressed parsimoniously in matrix form as:

\begin{equation}
\mbox{Attention} (Q,K,V) = \mbox{softmax}\left(\frac{1}{\sqrt{d_k}}\mathbf{Q}\mathbf{K}^T \right)\mathbf{V},
\label{Eq: Attention}
\end{equation}
where $\mbox{Attention} (Q,K,V) = [\mathbf{y}_1,\mathbf{y}_2,\dots,\mathbf{y}_T]$, the output vector. 

The product $\mathbf{Q}\mathbf{K}^T$ is usually referred to as the similarity matrix between the query and the key. From the dimensions of the $\textbf{Q}, \textbf{K}$ matrices, the dimensions of this product are $T\times T$. Finally, from Eq. \eqref{Eq: Attention}, the output dimensions of $A(Q,K,V)$ are $T\times d_v$ (remember that we initially started with an input matrix $T\times d_e$).


\begin{figure*}[!h]
\centering
\includegraphics[width=0.9\textwidth]{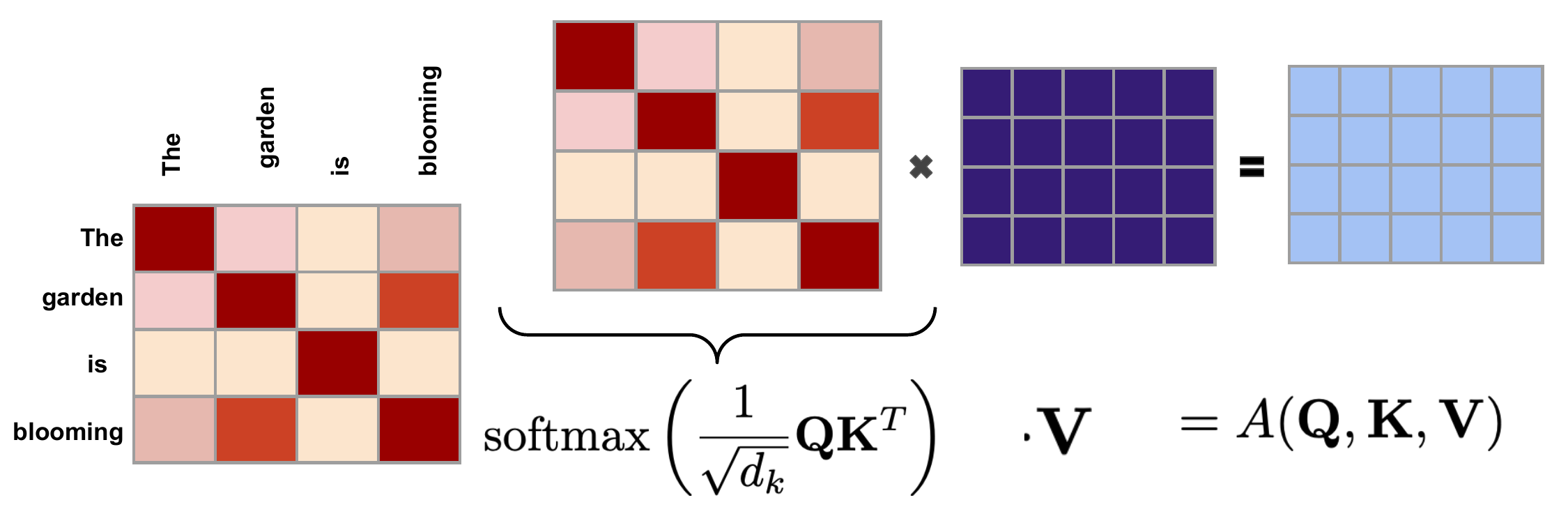}
\caption{The attention weights capture the similarity/relevance between the input vectors (left). For example, the words ``garden" and ``blooming" are more strongly correlated. Then these weights can be used to extract features that can be subsequently passed to a deep network. Adapted with modifications from
\url{http://introtodeeplearning.com}.
}
\label{fig: Attention_Feature}
\end{figure*}

To summarize, the similarity matrix $\mathbf{Q}\mathbf{K}^T$ is used to create the attention weights that capture the relevance of the combination of inputs (their correlations). In Fig. \ref{fig: Attention_Feature} we show an NLP task where the inputs are words and the attention weights correspond to the most informative word combinations. Then these attention weight matrices  multiply the matrix of value vectors to \textit{extract} important features that are subsequently passed to the next stage of the network.   
Admittedly, the introduction of the query, key, and value  matrices appears arbitrary if one begins with the introduction in the previous sub-section. The choice of three learnable projection matrices however strikes a balance between complexity and performance. The blogs listed below attempt to give a qualitative understanding of the three projection matrices, or you can query ChatGPT itself to explore how the attention mechanism works. 

\subsection{Multi-head attention}

\begin{figure}[!ht]
\centering
\includegraphics[width=1.1\textwidth]{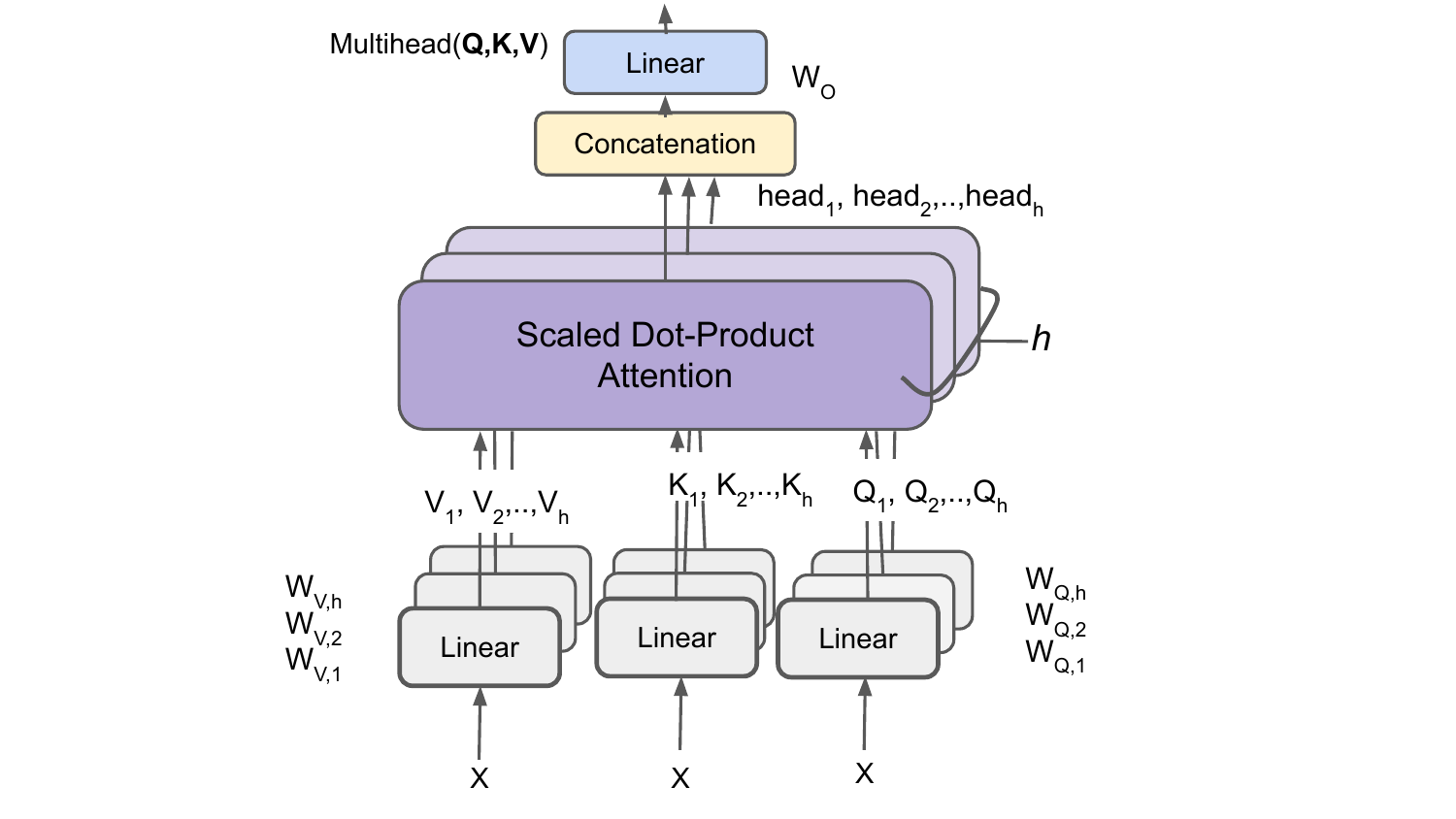}
\caption{Multi-head attention with $h$ attention heads. Adapted with modifications from Vaswani et al (2017),  arXiv:1706.03762. } 
\label{Fig: Multi_Head}
\end{figure}

We have introduced the learned projection  matrices $(\textbf{W}_Q,\textbf{W}_K,\textbf{W}_V)$. Combined, these are known as an \textit{attention head}. These matrices learn useful relations between the input vectors (e.g. between words in a sentence or parts of an image).  
Similar to the case of CNNs, where one introduces multiple filters (learned weight kernels) to learn different types of features from an image, we can introduce multiple (learnable) attention heads that can capture different relationships between the input tokens. 

To write down compact expressions for multi-head attention, let's note that the operations described in the previous section can be repeated $h$ times, \textit{in parallel}, producing matrices $\mathbf{Q}_i,\mathbf{K}_i,\mathbf{V}_i, \,\, i=1,\dots,h$, as shown in Fig. \ref{Fig: Multi_Head}. The multi-head attention output is then given by:

\begin{equation}
    \mbox{MultiHead}(\mathbf{Q},\mathbf{K},\mathbf{V}) = \mbox{Concat}(\mbox{head}_1,\dots,\mbox{head}_h)\mathbf{W}^O,
\label{Eq: MultiHead}
\end{equation}
where:
\begin{equation}
\mbox{head}_i= \mbox{Attention}(\mathbf{Q}_i,\mathbf{K}_i, \mathbf{V}_i)
\end{equation}
and $\mathbf{W}^O \in \mathbb{R}^{d \times hd_v}$ is a linear projection matrix that is used in order to project the concatenated vector to the output model space (i.e. it gives the output array of vectors $\textbf{y}_i$ that can be subsequently passed through feed-forward layers etc). 

It is worth paying attention to the input and output dimensions in Eq. \eqref{Eq: MultiHead}. As always, the input is a sequence of $T$ vectors of embedding dimension $d_e$ (we can represent the input as a matrix of shape $T\times d_e$). From our discussion in Sec. \ref{Sec: QKV_parametrization}, the dimensions of $h_i=\mbox{Attention}(\mathbf{Q}_i,\mathbf{K}_i, \mathbf{V}_i)$ are $T\times d_v$. Thus, we have that:

\begin{equation}
[\mbox{MultiHead}(\mathbf{Q},\mathbf{K},\mathbf{V})] =\underbrace{ \underbrace{\mbox{Concat}(\overbrace{\mbox{head}_1}^{T\times d_v},\dots,\mbox{head}_h)}_{T\times (d_v\cdot h)}\overbrace{\mathbf{W}^O}^{(d_v\cdot h)\times d_O}}_{T\times d_O}.
\end{equation}

The input matrix of dimensions $T\times d_e$ has been transformed into a matrix of dimensions $T\times d_O$; the initial embeddings have been transformed to new ones, that capture informative correlations between the input vectors to improve on the performance of a single attention head.

\subsection{Positional Encoding}

The self-attention mechanism, as we have described it so far, is oblivious to the position of the inputs ($\mathbf{x}$). In other words, unlike CNNs and RNNs, the Transformer architecture is invariant under permutations. However, for text sequences (sentences) or images, the order (or position) of the input token is important. A sentence will have a very different meaning (or lose any meaning!) depending on the order of words that compose it.  

In order to include positional information in the transformer architecture, one can add a vector $\mathbf{p}_i \in \mathbb{R}^{d}$, where $i$ denotes the position in the sequence, to the embedding vector $\mathbf{x}_i$ (with $d$ denoting the embedding dimension, previously denoted $d_e$):

\begin{equation}
 \mathbf{x}_i   \leftarrow \mathbf{x}_i + \mathbf{p}_i.
\end{equation}


The elements of the position vector $\mathbf{p}_i$ can be learned (in that case it is called a positional \textit{embedding}) or set to have a specific form that \textit{encodes} the position. Note that a different encoding may be needed for data that is both sequential and spatial, e.g. to capture the temporal and spatial positions of climate data. In the original attention paper, a sinusoidal \textit{positional encoding} vector was proposed: 

\begin{equation} 
    \begin{cases}
    \mathbf{p}_i(2j+1) = \cos \left(\frac{i}{10000^{2j/d}} \right),\\ 
    \\
    \mathbf{p}_i(2j) = \sin \left(\frac{i}{10000^{2j/d}} \right),
    \end{cases}
\end{equation}
For $j \in \{0,1,\dots,\lfloor d/2 \rfloor\}$ denoting the elements of the vector, where $\lfloor x \rfloor$ is the floor function, denoting the greatest integer that is not greater than $x$. Why choose this unusual form for the positional encoding, rather than an integer that exactly represents the position, or a monotonic function of it? The sine and cosine functions were chosen because they output a fixed range, regardless of the dimension $d$. The different ``frequencies'', the denominator of the argument, scale as $j/d$ to ensure that a unique position is associated with each token. That said, the particular functions given above are the choice made by Vaswani et al. for positional encoding and other choices are certainly possible. 

More recent transformer architectures (such as the GPT family) use learned positional embeddings, meaning that the network itself is free to find the optimal vector elements that encode positional information instead of imposing a specific functional form.

\section{The transformer block}

\begin{figure}[!ht]
\centering
\includegraphics[width=0.85\textwidth]{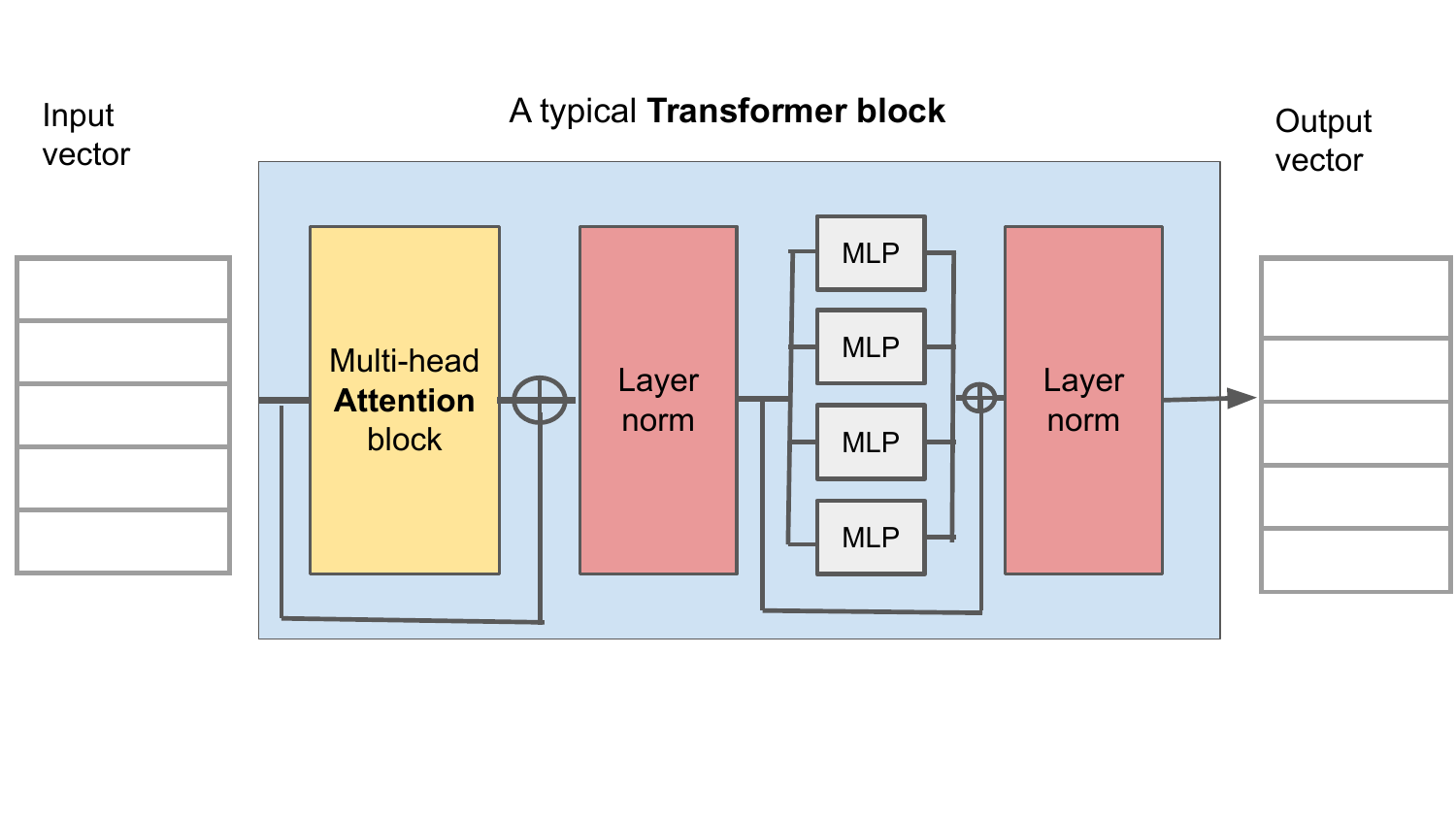}
\vspace{-0.4in}
\caption{Graphical representation showing the main elements of a transformer block. Adapted with modifications from \url{https://peterbloem.nl/blog/transformers}. } 
\label{Fig: Transformer_Block}
\end{figure}

We have casually mentioned transformers a few times so far, but what actually is a transformer? Although initially the term transformer was a reference to the architecture proposed in Vaswani et al (2017) for the task of machine translation, it is now customary to say a \textbf{transformer is any architecture that has a transformer block as a basic building element}. A transformer block  in turn implements the multi-head self-attention we  described above (note that other mechanisms beyond the scaled dot-product self-attention exist, but their description is beyond the scope of these notes). This section gives a brief overview of the transformer block and assumes some familiarity with deep learning. 

We can see a graphical representation of a typical transformer block  in Fig. \ref{Fig: Transformer_Block}. It is composed of a self-attention layer, followed by \textit{layer normalization}, a series of standard multi-layer perceptrons (MLPs) forming a feed-forward layer, and one more layer normalization. There are also some skip connections (gray lines that go under the Multi-head Attention block and the MLPs. Let's discuss the elements of a transformer block in a bit more detail. 

\begin{itemize}
    \item Skip connections: Similar to the skip connections that one can find in residual neural networks (ResNets), the skip connection adds the input $\mathbf{X}$ to the output of the multi-head attention. If we denote as $\mathbf{Z} = \mbox{MultiHead}(\mathbf{Q},\mathbf{K},\mathbf{V})$, then after the skip connection:
    \begin{equation}
        \textbf{Z}' = \textbf{Z} + \textbf{X}.
    \end{equation}
where $\mathbf{X}$ is a matrix containing the input vectors (including the positional encodings). This helps retain some  information about the original sequence.

 \item Layer normalization: Layer normalization is a variation of the batch normalization  seen in other  deep learning architectures for performance optimization. Let's assume a hidden layer with $H$ hidden units, $h_i$. Layer normalization standardizes those hidden units to have zero mean and unit variance by first calculating the mean and variance over the $H$ units:
 \begin{equation}
    \mu = \frac{1}{H}\sum_{i=1}^H h_i,
\end{equation}
\begin{equation}
    \sigma = \sqrt{\frac{1}{H}\sum_{i=1}^H(h_i-\mu)^2}.
\end{equation}
And then redefining the output of every hidden unit as:
\begin{equation}
\hat{h}_i = \gamma \frac{(h_i - \mu)}{\sigma},
\end{equation}
with $\gamma$ being a trainable parameter.  Layer normalization improves both the training time and the generalization performance of the transformer.

  \item Multi-layer perceptron (MLP): A single MLP, a feed-forward neural net, is applied independently to each output vector after  layer normalization. Adding this layer increases the capability of the model without increasing its computational complexity because the MLP acts on each position independently (the attention mechanism having already learned the correlations across positions). The result of this additional nonlinearity in the model is that the transformer is able to learn more about the patterns and relationships in the data. 

  \end{itemize}

  That's it! We have covered all the main elements of the transformer block. You should be able to understand the construction of more complicated transformer-based architectures and models, like the original transformer encoder-decoder model, the GPT family etc.  In the FAQs, we point out a good starting point to implement your own transformer. We have not discussed topics including pre-training, self-supervised vs. unsupervised training, computational issues, and  others that have emerged in the use of transformers. Some of these issues will be briefly discussed in the next section for astronomical applications and others in the FAQs in the following section. 

\section{Transformers in astronomy}

Sequential data in time domain astronomy is the numerical counterpart to NLP and has been tackled using transformers in several recent studies, as discussed in the first subsection below.  Image analysis problems with  large datasets occur in a variety of settings in astronomy and cosmology. The second subsection  summarizes the applications of ViTs for image analysis and concludes with multi-modal datasets in astronomy. This section assumes some familiarity with both astronomical surveys and deep learning. 

\subsection{Time series data}
Time-varying light sources, such as astronomical transients and variables, are often described in terms of their \textit{light curves}, or brightness over time. These time series are a natural application for sequence modeling architectures like transformers, but the prevalence of measurement noise and irregular sampling makes for a unique challenge for these text-focused models. A number of recent works have explored these applications, and can broadly be categorized in terms of their architectures and training procedures.

Most of the existing literature uses transformers in a supervised learning paradigm, where the attention mechanism and transformer blocks are used to produce learned features from the input sequences, which are then used to train a feed-forward classifier. AstroNet \cite{astronet} achieved impressive results on classifying 14 types of astronomical time series with such an architecture and tackled the irregular sampling issue with Gaussian process interpolation and resampling at regular intervals. ATAT (Astronomical Transformer for time series and Tabular Data, \cite{ATAT}) was tested on 32 object types and incorporates light curve metadata, such as redshifts, into its transformer-based architecture. It also uses a system of learnable Fourier coefficients to address the irregular sampling issue, bypassing the need for expensive Gaussian process regression. Inferring physical properties from light curves using transformers, such as surface gravity from stellar light curves or distinguishing false positives from exoplanet transit light curves, were shown by Astroconformer \cite{astroconformer} and \cite{Salinas_2023}. Finally, transformer-convolutional hybrid models have been successful at image-based tasks in astronomy, such as 'image time series' classification (ConvEntion \cite{convention} and galaxy morphology classification AstroFormer \cite{astroformer}).

A few works in astronomical time-series have leveraged the \textit{pretraining} paradigm, in which an initial training step is done with an intermediate self-supervised objective (e.g. masked data reconstruction, next token prediction) in order to learn transferable features that are reusable for a variety of downstream tasks. One such example is the TimeModAttn model, which uses an autoencoder architecture with attention mechanisms in the encoder to optimize a state-space model reconstruction of each light curve in the pretraining stage, then using the learned features for a downstream supernova classification task \cite{Pimentel_2022}. Following the masked language modeling objective introduced in BERT \cite{bert}, Astromer used a masked reconstruction loss to pretrain a transformer model for astrophysical variable objects and demonstrated good downstream classification performance on objects from multiple surveys \cite{Donoso_Oliva_2023}. An investigation into the optimal positional encoding (PE) paradigm for the unique properties of light curves was undertaken by \cite{morenocartagena2023positional}, showing that trainable PEs produced the best downstream results on classification of variable stars.

Finally, pretraining does not have to be purely in service of downstream tasks; pretrained models can be used for their generative capabilities alone. A denoising time-series transformer was introduced by \cite{denoising} to extract better signal from exoplanet transit light curves through masked reconstruction.

\subsection{Image analysis and other  applications in astronomy}

\begin{figure}[!ht]
\centering
\includegraphics[width=0.85\textwidth]{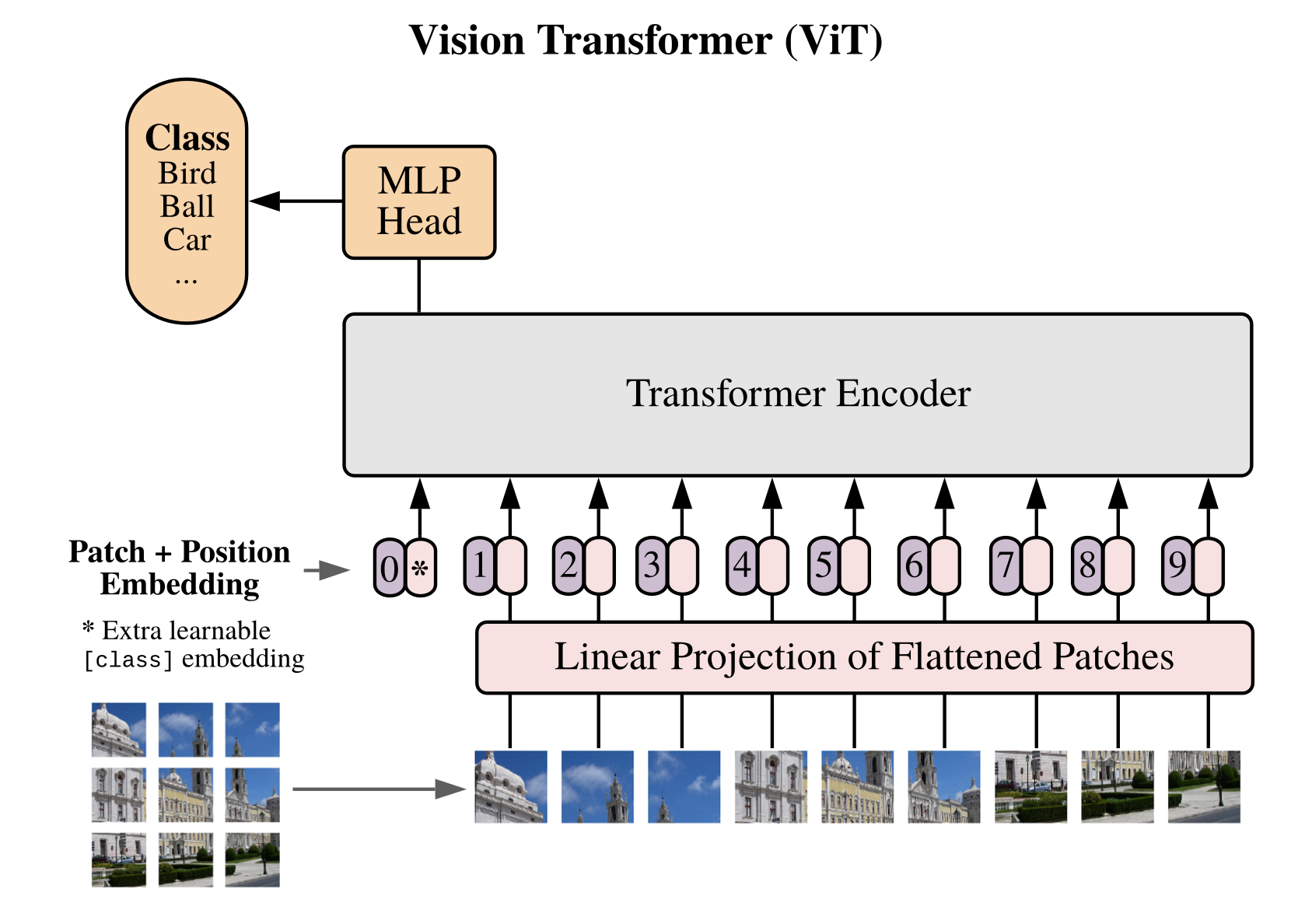}
\vspace{0.1in}
\caption{Graphical representation of Vision Transformers from Dosovitskiy et al (2021), arXiv:2010.11929. 
}
\label{Fig: ViT}
\end{figure}

Recent reviews of machine learning in astronomy and cosmology highlight the breadth of applications beyond time series data \cite{dvorkin2022machine,Huertas_Company_2023,Moriwaki_2023}. 
CNNs have been the dominant deep learning methodology to analyze images in astronomy (as in other fields). They have been used for for both classification and regression tasks, ranging from galaxy classification, cosmological analysis of galaxy and dark matter maps, to strong lensing and other applications. 

Vision Transformers (ViTs) have shown promise in image analysis tasks provided a suitably large training dataset is used. 
In order to convert a two dimensional pixelized image into sequential data, a ViT splits the image into patches (containing many pixels each) and  flattens the image into a one dimensional array of the patches. This array is then fed to a transformer encoder (typically after projecting into a lower dimensional vector space). This is shown in Figure \ref{Fig: ViT} from the paper introducing ViTs by \cite{dosovitskiy2021image}. 
ViTs perform well when pre-trained on large dataset (order $10^{7-8}$ images) and then fine tuned with training data for the application at hand. They are able to learn long range dependencies (correlated features across the flattened image) in the early layers of the network, unlike CNNs which apply local filters to the image and require several convolution layers to learn these. See the FAQs for more on the comparison of the performance of ViTs vs. CNNs. 

In astronomy, a handful of studies have applied ViTs to object detection, galaxy image classification and cosmological inference problems. \cite{merz2023detection}  used transformers for the simultaneous tasks of object identification, deblending, and classification on coadded image  from the Subaru Telescope. \cite{Jia_2022} appleid ViTs for detection of strongly lensed cluster arcs. As noted above, a combination of CNNs and ViTs has been used for image time series classification and galaxy morphology classification [5,6].  \cite{hwang2023universe} 
compared CNNs and ViTs for cosmology inference using 3D maps of halos (designed to represent galaxies) drawn from lightcone data generated by N-body simulations. The halo positions were recorded in $64^3$ voxels and fed to CNNs and ViTs to obtain constraints on  key cosmological parameters. 

Finally, \cite{leung2023astronomical} 
have made an attempt at building a foundation model for stars, and \cite{różański2023spectral} 
for spectral fitting, perhaps precursors to similar attempts for other branches of astronomy (see \cite{bommasani2022opportunities} for discussion of foundation models). Several  studies using transformers for analysis of images or other astronomical datasets are currently underway so the brief summary of papers above is no more than a snapshot of a growing branch of machine learning in astronomy.


\section{Frequently Asked Questions}
\subsection{FAQ: LLMs}
\begin{itemize}
    \item Q: Is the description of self-attention given in Eqns. 1-13  applied to each prompt? \\
    Yes. Say you ask a question using 7 words. (We'll use ``words" and ``tokens" interchangeably here for simplicity.) These input tokens are embedded in a vector (called $ \mathbf{x}$ in Section 2 above). This vector is a set of real numbers, its components  in some appropriate number of embedding dimensions, say 15. That's the starting point for the query key value operations.  
    
    \item Q: So is the output really produced one word at a time? How does it generate pages of coherent text? \\
    One word at a time is right. The first word generated is then used as the last token of the next input sequence to generate the second word. And so on! This is referred to as the auto-regressive strategy. 
    
    \item Q: How do the trillions of words used in the training data come into play in generating the response? \\
    The training data is used to fix the  free parameters of GPT's transformer architecture, mainly in the embedding vectors, the elements of the query, key and value matrices and the MLP layer. The total number of free parameters is also of order a trillion! A new prompt is run through the  pre-trained transformer to generate responses. 
    
    \item Q: Wow, that's a lot of training to answer my dumb questions. Surely the entire internet isn't thrown at the transformer in one go? \\
    Right, the training is done in ``mini-batches'', which could be a hundred words each. And remarkably, each mini-batch is run through the whole transformer to compute the loss by comparing the response (yes, one word at a time) to the ``truth''. The weights are updated using the usual methods of neural nets for all mini-batches, and this process is repeated until a stopping criterion is met.   
\end{itemize}

\subsection{FAQ: Astronomical applications}
\begin{itemize}
    \item  Q: For time series data, can I replace ``word'' or ``token'' with ``flux'' and infer the same things as above? \\
    Pretty much. Say you use a big synthetic dataset for self-supervised learning with transformers, and then generate new light curves. It will work the same way, putting out one flux value at a time. 
    \item Q: I'm convinced that transformers are powerful for sequential data like time series in astronomy. How about other applications in astronomy, like image analysis: will ViTs make CNNs history? \\
    Good question. The short answer is, No -- CNNs are still useful since the performance of ViTs is worse than CNNs until a certain threshold of training data is reached (approximately millions of examples [21]). Much of astronomy is operating in a low data regime, so CNNs may remain a better choice. Computational costs are the other big consideration in choosing the deep learning architecture. 
    
    \item Q: What are the prospects for interpretability when applying transformers to astronomical data? \\
    Interpretability or explainability of deep learning is a major challenge, and is probably responsible for its slower acceptance in fields like cosmology where alternative approaches that use (more) interpretable statistics are already well developed.  Attention maps and related tools have been developed for transformers to visualize the important elements of the input data for transformers. They may be more helpful than saliency maps for CNNs, which have well known issues with robustness, but these are early days -- definitely a research area in deep learning for all forms of data. 

    \item Q: What about a dataset/problem in astronomy makes a transformer suitable? \\
    Transformers are a natural choice for time-series or sequential data, though the noise and irregular sampling plaguing most astronomical data makes the naive application of transformers slightly challenging. For image data, ViTs generally outperform CNNs only after a certain threshold of training data as noted above, so ViTs may be a better choice when data is abundantly available.

    \item Q: Ok, I’m convinced to try a transformer approach to my data which I’ve previously analyzed with (RNN/CNN). How should I start? \\
    Great! \href{https://huggingface.co/}{Hugging Face} is a great resource with pretrained models (e.g. \href{https://huggingface.co/google/vit-base-patch16-224}{ViT pretrained on ImageNet}) as well as the code for popular models for training from scratch through the \href{https://huggingface.co/docs/transformers/index}{Transformers library}. Dataset preprocessing and loading can also be handled through \href{https://huggingface.co/docs/datasets/index}{Hugging Face Datasets}, which ensures a seamless full pipeline from data to results!
\end{itemize}

\textbf{\textit{Acknowledgements:}} We are grateful to Sudeep Bhatia, Mike Jarvis, Junhyong Kim and Colin Twomey for discussions and helpful feedback on these notes. 

\section{References and where to go from here}
\label{sec: References}
Far from being original, the description of transformers above is based on a number of excellent tutorials. We recommend the following resources for a deeper understanding of transformers, their historical development, and alternatives to the self-attention mechanism described here.
\\
\begin{itemize}
        
    \item   Vaswani A., Shazeer N., Parmar N., Uszkoreit J., Jones L., Gomez A.~N., Kaiser L., et al., 2017, arXiv:1706.03762.: \textit{The original ``Attention is all you need" paper that started the Transformer revolution.}
    
    \item   \url{https://peterbloem.nl/blog/transformers}: \textit{Excellent blog post; our treatment here was \textbf{heavily} based on it. Strongly recommended}
    
    \item   \url{https://sebastianraschka.com/blog/2023/self-attention-from-scratch.html}:  \textit{Blog post with code  to understand the attention mechanism.}  
    
    \item  \url{https://paperswithcode.com/paper/attention-mechanism-transformers-bert-and-gpt} \textit{Nice, technical, survey paper. Also covers the historical development of the attention mechanism, as a mechanism to improve RNNs, }
    
    \item 
 \url{http://introtodeeplearning.com/}: \textit{Excellent introduction to deep learning, including sequence models and attention.}
    
    \item  \url{https://lilianweng.github.io/posts/2018-06-24-attention/}: \textit{A great blog post by Lilian Weng. Discusses the evolution and different flavors of the attention mechanism in detail. }
    
    \item  \url{https://lilianweng.github.io/posts/2020-04-07-the-transformer-family/}: \textit{Very good discussion of different transformer types.}
    
    \item  Dosovitskiy A., Beyer L., Kolesnikov A., Weissenborn D., Zhai X., Unterthiner T., Dehghani M., et al., 2020, arXiv:2010.11929: \textit{The paper that introduced Vision Transformers (ViT).} 
    
    
    \item  \url{https://nlp.seas.harvard.edu/2018/04/03/attention.html}: \textit{Provides PyTorch code for each of the elements of the original Vaswani et al (2017) transformer paper }.
    
    \item  \url{https://uvadlc-notebooks.readthedocs.io/en/latest/tutorial_notebooks/tutorial6/Transformers_and_MHAttention.html}
\end{itemize}

\section{Astronomy References:}
\nocite{*}
\printbibliography[heading=none]

\end{document}